\documentstyle[11pt,newpasp,twoside,epsf]{article}
\begin{document}
\title{The Missing CV Population: Results From An Objective Prism Survey}

\author{C. Tappert}
\affil{Grupo de Astronom\'{\i}a, Departamento de F\'{\i}sica y Matem\'aticas, 
Universidad de Concepci\'on, Casilla 160-C, Concepci\'on, Chile}

\author{T. Augusteijn}
\affil{Isaac Newton Group, Apartado 321, E-38700 Santa Cruz de La Palma, Canary 
Islands, Spain}

\author{J. Maza}
\affil{Departamento de Astronom\'{\i}a, Universidad de Chile, Casilla 36-D, 
Santiago, Chile}

\begin{abstract}
We present the results of a search for cataclysmic variables (CVs) in the
Cal\'an-Tololo survey. We detected a total number of 21 CVs, 12 of them are 
previously unknown objects. Our results suggest that the mismatch between the 
theoretically predicted sample and the observed one is not due to observational
bias but has to be resolved by a revision of the theoretical models.
\end{abstract}

\section{The survey}
Theoretical models of the CV population predict that the vast majority consists
of low-mass-transfer systems, 
which have passed a theoretical minimum period $P_{\rm min,the} \sim 65$ min 
and are evolving back to longer periods. Observationally, this should cause a 
pile-up of systems at $P_{\rm min,the}$ (Stehle, Kolb, \& Ritter 1997). 
However, the {\em observed} minimum period is $P_{\rm min,obs} \sim 78$ min, 
and neither the large number of evolved CVs nor the pile-up are seen (Patterson
1998). CVs of this `missing population' are expected to be intrinsically very 
faint and to have long outburst recurrence times, making them hard to discover.
The presently observed sample is likely to be biased towards bright, young CVs, 
and does not serve as an unambiguous testbed for theoretical models.

Spectrocopic surveys should be ideally suited to discover evolved 
CVs, as low mass-transfer-rate systems generally show strong emission 
lines. The Cal\'an-Tololo objective prism survey (hereafter CTS; Maza et al.\ 
1989) covers 5150 deg$^2$ of the southern sky with 
$|b| \ge 20^\circ$ down to $\sim$18.5 mag. About half of the plates have been 
examined, and 59 candidate CVs have been selected visually on the basis of 
their spectral appearance on the plates. Follow-up observations included 
medium-resolution spectroscopy, calibrated and time-series photometry for all 
previously unknown objects, and time-resolved spectroscopy for selected 
identified CVs.

\section{Results}

\begin{figure}
\plotfiddle{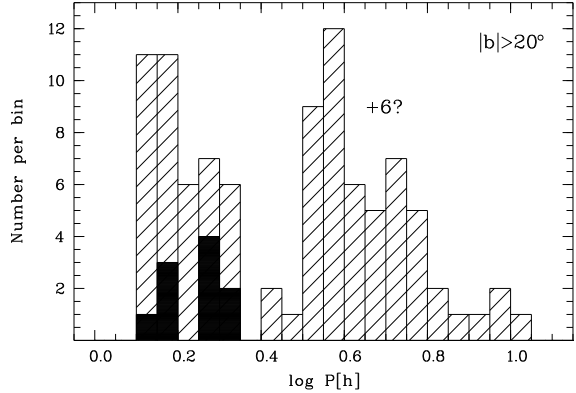}{4.41cm}{270}{29}{29}{-140}{155}
\caption{Period histogram of the non-magnetic CVs with $|b| > 20\deg$ in the 
Ritter \& Kolb (1998) catalogue (hashed) and in the CTS (solid). The number 
indicates the CTS CVs with unkown (and thus possibly, but not necessarily, long)
orbital period.}
\end{figure}

From the initial sample of 59 objects, 21 were identified as CVs (16 dwarf
novae and 5 magnetic CVs in low state), with 12 previously unknown 
objects. Figure 1 shows the period distribution of the CTS CVs in comparison 
with the known sample from the Ritter \& Kolb (1998) catalogue. Both samples
are limited to galactic latitudes $|b| > 20\deg$ and to nonmagnetic CVs, as 
the evolution for magnetic systems might be significantly different.
The restriction to high galactic latitudes does limit the bias towards distant,
long-period, high-mass-transfer systems, which affects the sample at low 
latitudes. As expected, our survey detected predominantly short-period CVs,
which are thought to be low-mass-transfer systems.

However, the number of newly discovered CVs is far too low to
solve the mismatch between theory and observations. Furthermore, although the
high-$|b|$ limited period distribution shows a (small) peak at 
$P_{\rm min,obs}$, the periods of the newly discovered CVs actually make this 
feature less pronounced (Fig.\ 1).

Unfortunately, the visual inspection of the CTS plates introduces a
different bias which is very difficult to quantify. It certainly affects to
a high degree the completeness of the sample, and also leads to a significant 
contamination by non-CVs. We remark however that this `personal' bias in 
principle should not be directed against evolved CVs. The fact that we did not 
find these systems therefore indicates that their number indeed is much lower 
than predicted by the theory, thus supporting the discussion and conclusions by
Patterson (1998).

\end{document}